# A New Covert Channel over Cellular Voice Channel in Smartphones

Bushra Aloraini, Daryl Johnson, Bill Stackpole, and Sumita Mishra
Golisano College of Computing and Information Sciences
Rochester Institute of Technology
Rochester, NY, USA

*Abstract*— Investigating network covert channels in smartphones has become increasingly important as smartphones have recently replaced the role of traditional computers. Smartphones are subject to traditional computer network covert channel techniques. Smartphones also introduce new sets of covert channel techniques as they add more capabilities and multiple network connections. This work presents a new network covert channel in smartphones. The research studies the ability to leak information from the smartphones' applications by reaching the cellular voice stream, and it examines the ability to employ the cellular voice channel to be a potential medium of information leakage through carrying modulated "speech-like" data covertly. To validate the theory, an Android software audio modem has been developed and it was able to leak data successfully through the cellular voice channel stream by carrying modulated data with a throughput of 13 bps with 0.018% BER. Moreover, Android security policies are investigated and broken in order to implement a user-mode rootkit that opens the voice channels by stealthily answering an incoming voice call. Multiple scenarios are conducted to verify the effectiveness of the proposed covert channel. This study identifies a new potential smartphone covert channel, and discusses some security vulnerabilities in Android OS that allow the use of this channel demonstrating the need to set countermeasures against this kind of breach.

## INTRODUCTION

Network covert channels represent a significant problem due to their security implications. Thus many research efforts have been focused on their identification, detection, and prevention. Covert channel identification is the process of discovering a shared resource that might be utilized for covert communication. This research contributes to the field by identifying a new network covert channel in smartphones.

Smartphones are always connected to the cellular network; however, little effort has been directed at investigating potential security threats with its covert communication. Previously, the cellular voice channel had never been used to launch such attacks to the best of our knowledge. This service was designed to carry audio only. Thus cellular service providers have not applied any information security protection systems, such as firewalls or intrusion detection systems, to guard cellular voice channel traffic in the cellular network core. Thus these channels are a prime choice over which to attempt a covert channel.

Theoretically, this channel could be employed in smartphones to conduct multiple covert malicious activities, such as sending commands, or even leaking information. As there are some past research that studied modulating data to be "speech-like" and transmitting it through a cellular voice channel using a GSM modem and a computer [1-3]. In addition to the fact that smartphone hardware designers introduced a new smartphone design that provides higher-quality audio and video performance and longer battery life [4-5], this research discovered that, the new design allows smartphone applications to reach the cellular voice stream. Thus information in the application could be intentionally or unintentionally leaked, or malware could be spread through the cellular voice stream. This could be accomplished by implementing a simple audio modem that is able modulate date to be "speech-like" and access the cellular voice stream to inject information to smartphones' cellular voice cannel.

This covert channel could be accompanied with rootkit that alters phone services to hide the covert communication channels. To investigate the potential threats with this covert channel, Android security mechanisms were tested and it was demonstrated that it is possible to build an Android persistent user-mode rootkit to intercept Android telephony API calls to answer incoming calls without the user or the system's knowledge. The developed modem along with the rootkit successfully leaked data from the smartphone's application and through cellular voice channel stream by carrying modulated data with a throughput of 13 bps with 0.018% BER.

## I. LITERATURE REVIEW

The covert channel concept was first presented by Lampson in 1973 as a communication channel that was neither designed nor intended for carrying information. [6]. A covert channel utilizes mechanisms that are not intended for communication purposes, thereby violating the network's security policy [7]. Three key conditions were introduced that help in the emergence of a covert channel: 1) a global shared resource between the sender and the receiver must be present, 2) the ability to alter the shared resource, and 3) a way to accomplish synchronization between the sender and the receiver [8]. The cellular voice channel has all three conditions, making it an ideal channel for implementing a covert channel. Network covert channel field research currently focuses on exploiting weaknesses in common Internet protocols such as TCP/IP [9], HTTP [10], VoIP [11], & SSH [12] to embed a covert communication. In the cellular network field, it has been demonstrated that high capacity covert channels in SMS can be embedded and used as a data exfiltration channel by composing the SMS in Protocol Description Unit (PDU) mode [13]. In [14] the authors introduced stenographic algorithms to hide data in the context of MMS to be used in on-time password and key communication. Cellular voice channel in smartphones has never been attempted before to the best of our knowledge.

BACKGROUND INFORMATION

### A. Smartphone Architecture

Smartphones consist of two main processors, the baseband processor (BP) and the application processor (AP). AP is responsible for the user interface and applications. BP has a Real-Time Operating System (RTOS), such as Nucleus and ThreadX, while AP is controlled by Smartphone OS such as Android. BP handles radio access to the cellular network, and provides communication protocols such as GSM, GPRS, UMTS, etc. AP is responsible for the user interface and applications. These two processors communicate through shared-memory or a dedicated serial channel. Cellular voice call routing and control are achieved typically by the BP only, whereas the AP handles all other multimedia functions. Thus, end users and applications are not able to access the cellular voice stream. However, as users spend more time using mobile phones for more media-rich applications, most smartphone hardware designers wanted to provide higher-quality audio and video performance and longer battery life. Therefore, they first introduced a separate dedicated processor for audio/video decoding to meet these increasing needs [4]. Then, hardware designers merged the audio digital signal processor (DSP) into the AP [5].

As has been discovered in this research, this new design has resulted in the audio routing functionalities, including cellular voice calling, being controlled by the AP (Figure 1) adapted from [4]. This feature introduced a new security vulnerability. The audio path to the cellular voice channel could be reached and controlled from the AP and potentially the end user.

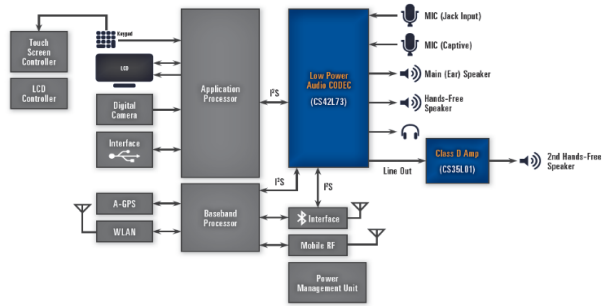

*Figure 1  Cirrus Logic Audio Subsystem Architecture show that the DSP, adapted from [4]*

### B. Cellular Voice traffic Overview

In digital cellular systems, like GSM and CDMA, when someone makes a phone call, the voice first passes to the microphone. It would then pass through the analog-to-digital converter (ADC) that converts the analog stream into digital data using Pulse Code Modulation (PCM) to be understood by the cell phone. The PCM method is utilized to represent sampled analog signals in digital form by recording a binary representation of the magnitude of the audio sample, and then the sample is encoded as an integer. The data stream then is processed and transmitted through the cellular core network in digital form.

The audio data stream, traveling among the cellular channels, is compressed to allow greater channel capacity. The cellular channels are band-limited channels, so audio signals should have frequencies within the telephone voice band which is between 300 and 3400 Hz. In order to reduce the bandwidth of the voice call and save the power of the cellphone, Discontinuous Transmission (DTX) is used to allow the cell phone transmitter to be turned off when the user is not talking. To do so and detect silence, DTX uses Voice Activity Detection (VAD) which is a unit that determines whether the speech frame includes speech or a speech pause to reduce the transmission to only speech and to reject noise and silence. Once the frame has been labeled as non-speech, it is dismissed instead of being transmitted. When the data stream is received on the other side, it is restored to the original source signal format and the digital–to-analog (DAC) module converts the bit stream back to audio wave.

### C. Android OS Overview

Android is an open source operating system based on the Linux kernel. The Android operating system consists of five software components within four main layers: Linux kernel, libraries, Android runtime, application framework, and applications (Figure 2). The first layer is the Linux kernel layer that provides main system functionalities, such as networking and device driver management as it communicates with the hardware and the BP.

The second layer includes two components: a set of libraries to provide multiple system services, such playing and recording audio and video, and an Android Runtime environment which has a main component, Dalvik Virtual Machine (DVM). DVM allows every Android application to run in its own process under a unique UNIX UID, and it is responsible for executing binaries of all applications located in the application layer. The third layer is the application framework that offers Application Programming Interfaces (API) to third party application developers. The fourth layer is the applications layer that is written fully in Java, and represents the installed user application. Applications are written in Java and compiled to the Dalvik Executable (DEX) byte-code format. Every application executes within its own instance of DVM interpreter.

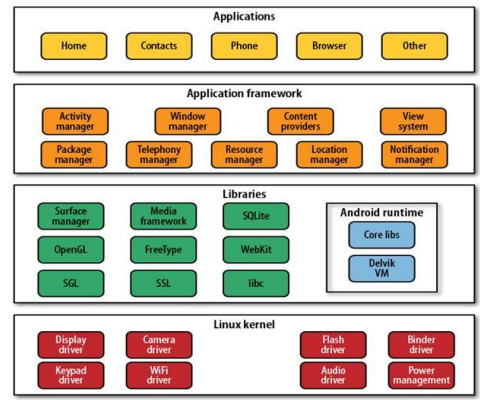

*Figure 2  Android Architecture Layer*

### D. Android Telephony Framework

Android telephony stack is responsible for the communication between BP and AP. Android telephony stack consists of four main layers: applications, framework, radio interface layer, and BP (Figure 3). The application layer involves all the smartphone telephony applications, such as Dialer, SMS, etc. Phone applications in the application layer communicate directly with the internal API in the telephony framework to place and tear down phone calls. Android telephony framework provides APIs for the phone application; however, APIs cannot be entered from any other applications that are not part of the Android system. The telephony requests are passed to the BP. BP replies to the application through the Telephony framework.

The communications between the telephony framework and the BP are handled by the Radio Interface Layer (RIL). The RIL communicates with the BP by utilizing a single serial line. RIL has two main components: a RIL Daemon and a Vendor RIL. The RIL Daemon connects the telephony framework to the Vendor RIL, initiates the modem, and reads the system properties to locate the proper library for the Vendor RIL. The Vendor RIL is the BP driver. There are various vendors, therefore, each vendor has a different implementation of the vendor RIL.

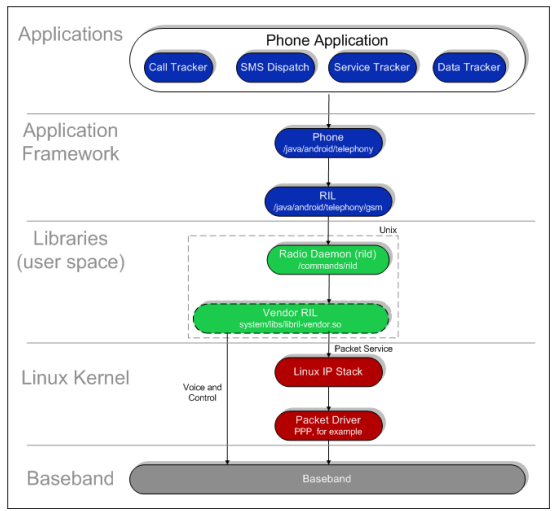

*Figure 3 Android Telephony Architecture*

### E. Android Media Framework

Since it is important to understand how smartphone cellular voice calls take place, it is essential to comprehend how Android handles audio streams which forms an important part in carrying out a cellular voice call. Android Application Framework takes care of Android's multimedia system; it uses the Android media APIs to call media native code to contact the audio hardware (Figure 4). The Android audio libraries include two native layers dealing with audio software: Audio Flinger and Audio Hardware Interface (HAL).

Audio Flinger communicates with the HAL layer which represents the hardware abstraction layer that hides audio drivers from the Android platform. Audio Flinger is the audio server that provides some required audio functions, such as audio stream routing and mixing, since the audio stream can be either input or output to/from multiple microphones, speakers, and applications. Audio Flinger performs audio routing by setting a routing mode, such as MODE_IN_CALL, MODE_IN_COMMUNICATION, or MODE_NORMAL, which is then passed into audio.h interface in Audio HAL which, in turn, determines the routing path.

When some applications use Audio Flinger to redirect the audio stream as STREAM_VOICE_CALL, some vendor-specific audio HAL implementations redirect the stream that comes from/to an application from/to the actual stream voice call by default. In other smartphones, redirecting voice call stream from/to an application from/to the actual cellular voice stream is also possible, since the audio routing is currently accomplished using the application processor, not the BP as mentioned earlier. This is achieved by making some modifications to audio HAL and some application framework components, such as Audio Service and Audio System in smartphones that cannot redirect the required audio path by default.

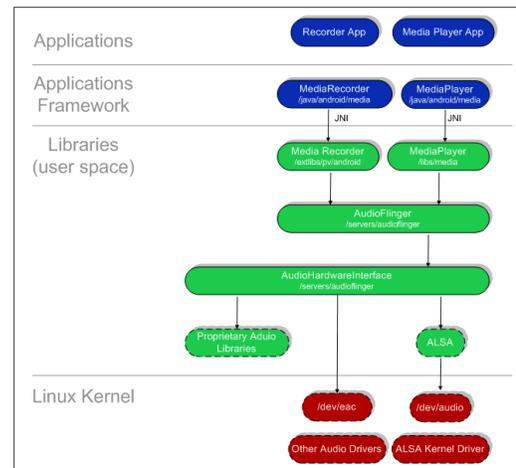

*Figure 4 Android Audio System*

### F. Android Phone Calls

In Android, phone calls are achieved using both the telephony and media frameworks. The phone application runs inside the com.android.phone process, which is comprised of multiple components such as SIP, SMS, and phone application. All of these components utilize the telephony framework APIs to communicate to the BP through an RIL socket. Hence, the phone application communicates to the BP by calling the framework APIs to initiate and receive voice calls. When an incoming call reaches the destination cell phone, BP sends a request to RIL, which communicates to the phone application as a ringing call. Once the user answers the call, the phone application sends a message to the RIL to open the voice channel, and the Android media system switches on the voice call channel.

### G. Android Security

Android framework involves multiple security mechanisms, such as application sandboxing, application signing, and Android permission framework. The application sandboxing mechanism is employed to isolate an application from system resources, actual hardware, and from the other application resources. Sandboxing is performed by assigning a unique user ID to each application to ensure that it has its own sandboxed execution environment. Application signing is a key security mechanism in Android. All Android applications must be signed independently with a self-created key using public key infrastructure. Applications that are signed with the same key could share the same UID and play in the same sandbox.

The Android permission framework is utilized to control access to sensitive system resources by applying Mandatory Access Control (MAC) on Inter Component Communication (ICC). Therefore, each application defines a list of permissions that are approved during the installation process.

## MATERIALS AND METHODS

### H. Audio Modem Design

The modem design consists of an encoder and decoder pair which works entirely in software. The encoder converts raw data into "speech-like" waveforms to be injected into the microphone of the downlink voice stream. The audio signals then travel among the cellular voice channel as speech in digital form, and reach the second cell phone to be played over uplink voice stream's speaker. They are then demodulated and converted back into the original data. The designed modem utilizes Morse code combined with Frequency Shift-Keying modulation (Figure 5).

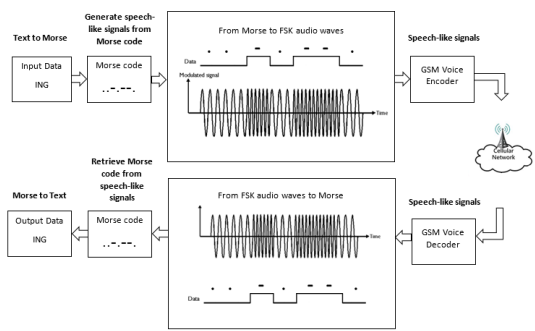

*Figure 5 Audio Modem Design*

### I. The Encoder

The encoder encodes text information into Morse code as a series of dots and dashes. Then it generates pure sine waves of two frequencies using FSK modulation: one frequency represents a dot and a different frequency represents a dash using the following equation for each sample in the buffer:

$$f(x) = Amplitude * sin(2\pi f) * (sample / sampleRate) \quad (1)$$

Where:
- Amplitude: the amplitude of the wave. It is a constant value for all waves.
- f: the frequency of the wave. This is a variable and it depends on if the current code is a dot or a dash.
- Sample: the current sample in the buffer.
- SampleRate: the modem sampling rate which is 8000 KHz.
- Sample / sampleRate: represents the current time.

*The audio modem works as following:*
1) Map each character to its corresponding Morse code.
2) Play a hail tone to indicate the beginning of the audio data.
3) For every code:
   - Play a tone with high frequency if the code is a dash.
   - Play a tone with low frequency if the code is a dot.
4) Play a silence to separate each character.
5) Play a long silence to separate each word.
6) Form each code to be a "speech-like" waveform of 200 ms frame length.

The frequencies that were chosen with the modem are 600 and 1000 to satisfy the demands of the cellular network voice channel bandwidth (300Hz and 3400Hz). In addition, the sampling rate was selected to be 8000 kHz to suit the telephony system sampling rate that carries speech signals. Each 200 ms voice frame is injected into the GSM encoder to avoid the distortion that could be issued from the VAD.

### J. The Decoder

The decoder does the reverse function of the encoder. The decoder performs spectral analysis to analyze each received sample and estimates its frequency to retrieve the original data. Fast Fourier Transform (FFT) algorithm was used to analyze the spectrum and detect the frequency pitch. Therefore, when the decoder receives an audio wave sample, the following steps are accomplished:

1) Low-pass filter is applied to the audio signal to get rid of noise and enhance the reliability of the pitch detection.
2) Hann window function is applied to the input data to increase detection quality by examining the important details of the signal rather than analyzing the entire signal, and to avoid signal leakage in both sides of the wave.
3) FFT is performed on the input data, which is an algorithm that quickly computes the frequency of a given signal. The output of the FFT is an array of N complex numbers that has real and imaginary parts.
4) Normalize the FFT bins to bring the peak amplitude to a normal level, and hence making all the received audio signal the same volume so as not to affect the quality of frequency detection.
5) Calculate the magnitude values of each FFT bin, using the following formula:

$$magnitude[i] = sqrt(real[i] * real[i] + imag[i] * imag[i]) \quad (2)$$

6) Get the bin of the maximum magnitude value in the FFT data.
7) Calculate the frequency of the index of the maximum magnitude value to detect the signal frequency. The frequency is calculated by this equation:

$$freq = i * SampleRate / N \quad (3)$$

Where:
- i = index of magnitude
- SampleRate = the modem sampling rate
- N = size of FFT data

8) Finally compare the received frequency to the dot and dash frequencies value with a minor tolerance.

### K. Modem Design Constraints and Challenges

To develop a modem that copes with cellular network channel characteristics, some constraints on audio signal needed to be taken into consideration.
1) The generated audio frequencies must fit into bandwidth between 300Hz and 3400Hz. Otherwise, an ultrasonic signal is generated that can't be heard by the human ear and is not suitable for this channel.
2) The generated audio signals should pass through the voice channel without being omitted by VAD which is primarily used to stop frame transmission at silence.
3) Cellular networks employ speech compression codeces in order to save bandwidth and enhance sound quality. As a result, these speech compression techniques could significantly distort audio signals that are not "speech-like", so the generated signal must retain typical speech characteristics.

In addition to the constraints mentioned above, there are a number of challenges in implementing such a modem. One of the challenges of the cellular voice channel is that a voice call between two cell phones is achieved in real time. Hence, there are no retransmissions in case of a missing voice frame. Sometimes, a voice frame could not be received by the other cell phone due to radio interference or the voice frame could be dropped intentionally by the core network when frame stealing scenario happens. Frame stealing usually happens during calls where a user receives an incoming waiting call request and the network drops the current voice frame in the conversation to send a notification to the user. On the other hand, smartphone capabilities, such as the processor and battery life, as well as the quality of audio hardware varies among cell phones; for example one cell phone speaker that sounds clear and loud could be low or unclear in another, as there are no standardizations in cell phone speaker and microphone specifications.

### L. Audio Modem Implementation

The modem works as follows: When there is an ongoing cellular voice call, the receiver presses on the receive button to obtain the required data. At the sender side, a text message is encoded into Morse code, then an Integer array containing the dots and dash frequencies is composed. This array is sent to the encoder, sampled as sine waves to generate audio tones and injected into the cellular voice stream using AudioManager.STREAM_VOICE_CALL audio track stream.

At the other end, after Receive button was pressed, it constantly records the uplink call stream to detect the hail key which refers to the beginning of the real incoming audio data. The hail key has a different frequency from the dots and dashes. It is generated once at the encoder to indicate the beginning of the audio data to be recorded. If the key was found, the Decoder decodes the audio waves using FFT algorithm and other signal processing functions to process the incoming signals and detect the frequencies. Once the audio signal frequencies have been determined, the decoder creates an Integer array of the detected frequencies which is then sent to the Morse decoder to retrieve the original data.

### M. Rootkit Design

Besides an audio modem, it was necessary to employ another system that compromises smartphone operating systems and open cellular voice channels by stealthily answering voice calls and performing data exchange. This was accomplished by utilizing a rootkit that hides its presence in the operating system and quietly allows continuous privileged access.

### N. Bypassing Android Security Mechanisms

To help the rootkit to catch the incoming calls before being delivered to Phone app, it was necessary to find a way to allow the rootkit to reach the RIL socket to listen to the BP messages. RIL socket is a Linux socket that is owned by com.android.phone process only, and it cannot be entered from outside the phone process. Thus, building the rootkit inside the com.android.phone process to communicate to the socket was the selected method to secretly communicate to the BP and to intercept its messages.

Designing a rootkit to work inside the Phone process requires overcoming the Android application sandboxing mechanism. Sandboxing is achieved by assigning a unique User ID for each application so the other applications cannot access its data and resources unless they have shared User ID, process, and their certificate match. The rootkit takes advantages of "android:sharedUserId" and "android:process" attributes in the manifest to play inside the com.android.phone process sandbox. The android:sharedUserId attribute allows two applications to share the same User ID so it can reach the other's application data and resources. The android:process sets the process where all the application's components should run.

Setting sharedUserId and process attributes on the manifest allows the rootkit to run in the com.android.phone process; however, it also requires the rootkit's certificate to match the Phone App's certificate unless the signature verification process in Android is bypassed during the installation time.

The Phone App application is signed by the system platform certificate and runs in the shared system sandbox as a system user. Therefore, it has more privileges to access the system resources. Signing the rootkit with the platform signature was obtained with custom ROMs, such as Cyanogenmod, because they are using the default certificate which is known as it is released publically.

Another method also used in this research is bypassing the signature verification method in order to infect the rooted Android official ROMs. That was achieved by modifying the Services.jar. The Services.jar has multiple java classes. One of these classes is PackageManagerService class responsible for matching the signature of the application during the installation time. The method in the PackageManagerService is called CompareSignatures, a Boolean method that returns true if the signature match and false if not. The method was modified to always return true which means the signature is always matched. This approach could be used by an attacker with any rooted Android easily.

After overcoming the security obstacles, the rootkit was successfully constructed inside the phone process. Yet it needs to go further to intercept the telephony framework APIs calls and take control of answering the incoming voice calls. To discover a process to intercept the telephony APIs calls and control the system call flow, it was necessary to understand how telephony applications really work to understand their weaknesses and vulnerabilities. However, as there is no documentation about the telephony applications, it was essential to analyze the phone app and the internal telephony framework source code. Therefore, in the following section a discussion about Android application architecture and the Phone App incoming call data flow are provided in order to discover any vulnerability that helps in achieve the task.

*O. Android Application Architecture*

In this section, an overview of how an Android application generally works is highlighted. When an Android application is not running, and one of its components starts, the Android system initiates a new Linux process with a single main thread of execution for the application. Then, when another component of the same application also starts, it will run in the same process and the same main thread.

The main thread is essential because all the application's components will run in it by default. The main thread is responsible for user interface interactions and events handling. It also handles system calls to each application component. Android thread can associate with a Looper which runs a message loop for a thread to process different messages. A Looper contains a MessageQueue that contains a list of messages. The interaction with the message queue is performed by using a Handler class (Figure 6).

The Handler is associated to the Looper and its associated thread. Handler is responsible for handling messages within the MessageQueue. When a message is delivered to a Handler, a new message arrives in the MessageQueue. When a thread creates a new Handler, it binds the Handler to it and to its message queue. Then the Handler can send messages to the message queue and handle messages as they arrive from the message queue.

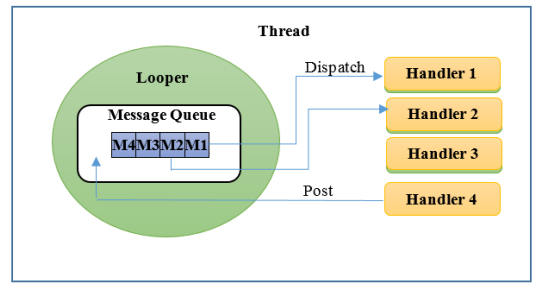

*Figure 6 Thread's Looper and Handler*

*P. Incoming Call Data flow*

When an incoming call request arrives to the BP, it is delivered to the Vendor RIL. The Vendor RIL passes the request to the RIL daemon which delivers the request to the telephony framework layer. The request message is passed among the telephony framework classes in this order: RIL, Basecommand, GSMCallTracker, GSMPhone, BasePhone and then CallManager. The CallManager communicates to the Phone App in the application layer of Android to show up as an incoming call (Figure 7).

Each class in the telephony framework can obtain notifications about the arriving messages by using RegistrantList object, which is simply an object that holds a Java ArrayList of Registrants object. So each class registers itself with the previous class's list in order to obtain notifications. The rootkit takes advantage of this design and registers itself with PhoneBase and receiveds the incoming notifications when they are delivered to the CallManager, and acts before CallManager Handler passes the incoming notification to the Phone App. Once the rootkit has received the incoming notification about a new incoming call, it obtains the incoming call's caller ID and compares it with a predefined number. If the number does not match, it leaves the notification to arrive at its final destination, which is the Phone App that shows up as an incoming call. However, if the caller ID matches the predefined number, it answers the call covertly.

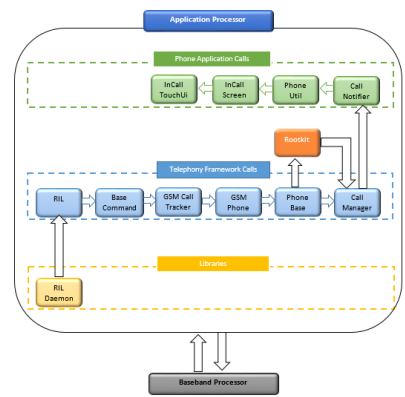

*Figure 7 An Incoming Voice Call Request Flow between the BP and AP in Android*

Android threads can only perform one task at a time. If a long running task has been executed in the Handler, the message queue cannot handle and process any other messages. In addition, it was discovered during this research that it is possible to conduct denial of service to the Looper, so the other process components could not receive new arriving messages. Android Looper is based on c code in native libraries. When trying to Send Message at Front of Queue twice manipulating the pointers, the message queue will be starved and the Looper hindered. This blocks the thread from delivering the new messages causing denial of service (DOS), or, in another scenario, changing the program flow in order to execute malicious code. Therefore, the message queue has been starved by sending two messages at the front of the message queue to hinder the Looper and blocks the thread from delivering the new messages. Thus, when the rootkit wants to answer the call covertly, it hinders the Looper from delivering the other messages until finishing the call. The Looper in the phone's main thread will not be able to deliver the other messages to the phone application. The rootkit acts as a proxy between the application layer and the telephony framework to transparently intercept and filter incoming voice calls from the BP (Figure 8). Whenever an incoming voice call request comes from the BP to the Phone App, the rootkit intercepts that incoming call request and decides whether to answer the call covertly in the case of calls from a specified number or to pass the call request to the Phone App to show the incoming call.

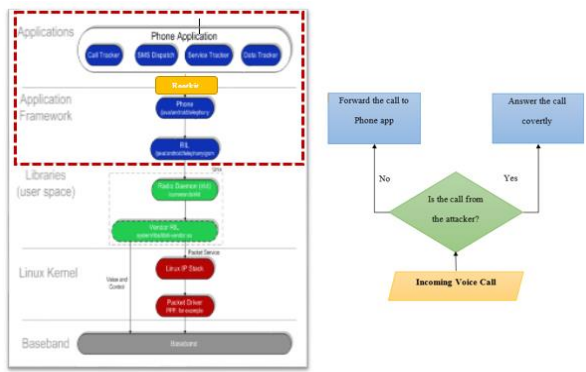

*(a) Rootkit Location in Android Telephony Framework*   *(b) Rootkit Data flow*

*Figure 8 Rootkit Location and Control Flow on Android's Telephony*

### Q. The Rootkit Implementation

The rootkit was implemented as an Android service that should be installed in the system partition of the application layer as it needs a privileged access. The rootkit works as follows: the rootkit service starts with the system boot and works as a persistent service, so when it is killed by the system due to shortage of the system resources, it restarts. When the service runs, it works inside the phone process and the phone process's main thread. The service creates a Handler and hooks itself and its Handler to the main Looper of the main thread in phone processes. Once the Handler has been created, it is able to obtain and send messages from/to the phone main thread's message queue.

When the Handler starts, it registers itself with the PhoneBase's Registrant List to listen to messages that come from the Phone Base object and to receive notifications before CallManager. When the Handler receives a new ringing connection message that comes from the Phone Base, it gets the caller ID and compares it with a predefined number. If the caller ID does not match, it leaves the message to arrive at its destination. If it matches, it answers the call before the Phone App has been notified and hinders the main Looper by starving its message queue. The Handler sends two messages to the front of the message queue hindering the main Looper.

In fact, while implementing the rootkit, it was essential to use the Android telephony internal classes and hidden components. However, as it is stated in Android documentation, these internal and hidden API telephony classes could not be reached by applications other than the phone process. This rule was simply achieved by excluding the jar files of the internal classes and the hidden APIs from the developer environment, such as Android Studio and Eclipse environments, and no extra steps were necessary to prevent using these APIs at the run time. The solution was simply to include these jar files in the developing environments in order to be able to use them.

### R. Rootkit Situations

The rootkit makes decisions based on the current situation of the phone call state as follows:
1) When a hacker calls during a call between the victim and another party, the rootkit will answer the hacker's call without being noticed by the victim.
2) When a hacker calls during an idle phone state, the rootkit will answer the call covertly and allow data exchange and spying.
3) When a victim tries to initiate an outgoing call during a hang-up call, or when a victim receives an incoming call request during a hang-up call by the hacker; the phone app will not respond to the request and will be idle because its main loop has been starved by a denial of service attack that was accomplished by the rootkit.

## EXPERIMENTAL DESIGN AND RESULTS

Using a real cellular infrastructure is more reliable than a simulated network; therefore, a real cellular network infrastructure and real smartphone were used during the experience to validate the theory. The T-Mobile GSM network was selected to conduct the experiment. The developed systems in this research could be run in all Android phones. However, in some Android devices, there is a need to modify some Android system audio files in order to control the voice call stream. The smartphones used in this experiment are Samsung Galaxy S3 and Galaxy Nexus. These two phones were chosen because of their ability to control the voice call stream by default; thus no extra work was needed to modify some system audio files to control the stream. The voice call was only used for the experiments. Both smartphones run Android Jelly Bean 4.3.3. Galaxy Nexus runs Cyanogenmod custom ROMs, and Samsung Galaxy S3 runs a rooted official ROM.

## S. First Scenario

The first scenario was implemented to investigate the ability to utilize the cellular network as a bidirectional covert channel. In this scenario, two individuals exchange text messages using a smartphone application and the cellular voice channel as a shared resource. In this scenario, only the audio modem was used and combined with a user interface to help the two individuals send and receive the messages. The application resides in the application layer. Thus, any kind of information could be leaked intentionally from the sender to the receiver.

**Test Results:** Figure 9 displays a screenshot of the smartphones to exhibit test results that were conducted during an active voice call. This successfully validates the modem's ability to utilize the cellular voice channel as a carrier of the generated audio signals by the smartphone. However, sometimes the receiver could not get the exact original data for several reasons, such as noisy environment, frame stealing scenario, smartphone audio hardware quality, and modem implementation.

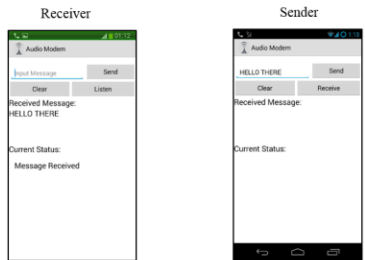

*Figure 9: The screens show a sender and a receiver forming a covert channel to exchange information as audio through voice call.*

## T. Second Scenario

The second scenario examines the ability to use the proposed covert channel to leak information unintentionally from one side while the other side desires a successful communication. In this scenario, rootkit works in the hacked smartphone and monitors and filters the incoming calls. Once rootkit has received an incoming voice call from a predefined number, it opens the voice channel covertly before it shows up in the smartphone's Phone App to allow data exchange. In this scenario, when the channel is opened, the last SMS in the hacked smartphone will be modulated and sent over the covert channel to be acquired by the other side. The other side uses the software audio modem application to modulate the audio waves and see the last SMS. In addition, theoretically any kind of data could be leaked, such as a picture or video; however, this implementation focuses on text data and was used only to validate the proposed covert channel.

**Test Results:** Figure 10 displays a screenshot of the smartphones to show test results. When the attacker made a call to the victim, rootkit recognized the attacker caller ID and, based on that fact, answered the call without showing up on the victim's screen. The victim had no idea about the ongoing voice call in his smartphone. Rootkit leaked the last received SMS in the victim's device by using the software audio modem. The attacker obtained the SMS by using the developed audio modem. However, the victim might hear audio waves played in his/her phone, but this issue can be overcome by using modulation techniques that simulate a natural sound like a bird or cricket, as these sounds could be played as a notification in some applications.

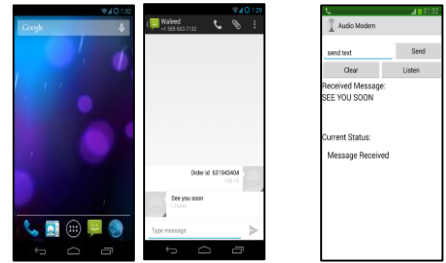

*Figure 10 The right screen shows when the attacker made a call to the victim, and in the left screen the rootkit in the hacked phone recognized the attacker's caller ID and based on that it answered the call without showing up on the victim's screen*

## U. Third Scenario

The third scenario was implemented to test the proposed covert channel to be involved in botnets as command and control C2 channel. With any botnet, C2 channel is the most significant part that contributes into its covertness and effectiveness. The designed system is similar to the second scenario as it combines both the audio modem and rootkit. However, when the rootkit opens the incoming voice channel based on a decision made regarding the caller ID number, the rootkit listens to any command that is sent by the other side and executes it directly instead of leaking information from the hacked device. Rootkit will act as a botnet that listens to commands and executes them as required. Table 1 includes some of the offered commands in this scenario:

Table 1 Some Commands with the Implemented Bot

| Command | Description |
|---|---|
| Reboot | Reboot the system. |
| Clrlog | Clear call log. |
| Blueto | Switch Bluetooth on. |

**weswaaaews**
**Test Results:** Figure 11 includes screenshots of the smartphones to show test results. When the attacker made a call to the victim, rootkit recognized the attacker caller ID and, based on that, answered the call without showing up on the victim's screen. Rootkit then waited to receive a command, and once it was obtained, executed it. The attacker sent the command using the developed audio modem.

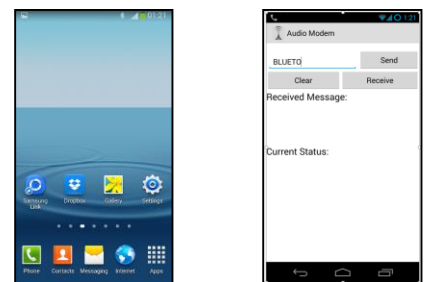

*Figure 11 The right screen shows the hacker's screen when he sent "Blueto" command to open the Bluetooth device in the hacked phone, the left screen shows the hacked phone when it respond to the command and turned on the Bluetooth device.*

## DISCUSSION

*V. Modem analysis*

In tested scenarios, when ideal conditions occurred where the surrounding environment is quiet, the smartphone hardware is loud and clear, air interface is not noisy, and the call is carried out over one speech compression technique. The results were perfect and accurate, as the second party got the exact sent message. In a realistic scenario, these conditions are not always guaranteed, so any conditions can easily either hinder the message from being transformed or omit some frames in the sent message. However, the audio modem design can be enhanced and optimized to overcome these constraints. The objective of the audio modem implementation is only to verify basic functionality and show possible scenarios that can be implemented successfully on real smartphones.

The modem implementation works in most Android smartphones, but the ability to access the voice call stream varies among Android smartphones. Some smartphones can reach the voice stream from Audio Manager directly; however, reaching the voice stream of a cellular call in other smartphones could be accomplished by modifying some Android system files. The current modem implementation with the capability to reach the voice call stream by default works on most Samsung Galaxy S series and Nexus smartphones.

*W. Rootkit Analysis*

User-mode rootkit design has two primary purposes—building a rootkit that is able to communicate to the BP and run covertly in any rooted Android OS, and filtering the incoming call to answer voice calls, if needed, before they show up on the Android screen. In addition, it was important to verify the ability to implement a portable rootkit that can work easily without modifying underlying system files or applications. In fact, the portability factor is significant to test the ability to deploy the rootkit in any smartphone device, which indicates the ease of its distribution among Android smartphones.

Rootkit implementation successfully verified what it was meant to do. Rootkit was able to be portable and work silently. The rootkit works in all Android-rooted stock ROMs, as well as most custom ROMs that have Jelly Bean 4.3.3 version or below, and it was not tested in newer versions. Rootkit was tested successfully in Samsung Galaxy S 3 I9300, Galaxy Nexus and Nexus S, Samsung Galaxy S 4, and Samsung Galaxy Duos y GSM versions, and is believed to work in most GSM and CDMA Android smartphones.

*X. Covert channel analysis*

The effectiveness of a covert channel can be evaluated by three factors; covertness, bandwidth, and robustness. Covertness determines to what extent the covert channel can be detected. The proposed covert channel is difficult to be detected because it is unknown by an adversary. Channel bandwidth refers to the channel maximum error-free transmission rate. Bandwidth is usually expressed in bits per second.

There are multiple factors effect on the cellular voice channel bandwidth. One factor is the used speech codec which affects the bit rate of the data. In addition, the speech codec can be switched during calls, because cellular network providers also can control and increase the number of active calls within one base station by switching cell phones to a low bitrate speech codec. That will impact the data transfer rate, as it could vary among the speech codecs and also in one speech codec at the time when the base station experiences an overload. In the proposed channel, the voice call throughput was entirely occupied in order to convey covert information, and it achieved a throughput of 13 bps with 0.018% BER.

The channel robustness determines the ease of limiting the channel capacity by adding noise or the ease of removing the covert channel. Therefore, the channel robustness is high, because it is not reliable to remove or add noise to this kind of channel; it will affect the legitimate data transferring quality.

## CONCLUSION AND FUTURE WORK

As smartphones are trending to increase their computational capabilities, employees and individuals increasingly rely on smartphones to perform their tasks, and as a result smartphone security becomes more significant than ever before. One of the most serious threats to information security, whether within organization or individual, is covert channels, because they could be employed to leak sensitive information, divert the ordinary use of a system, or coordinate attacks on a system. Therefore, identification of covert channels is considered an essential task. This research takes a step in this direction by identifying a potential covert channel which could affect smartphone security. It provides a proof of concept of the ability to use the cellular voice channel as a covert channel to leak information or distribute malware. It introduces details of designing and implementing the system and the challenges and constraints that have been faced to accomplish the system. It has been realized during this research that as smartphone hardware and software designs have changed recently, it allowed and contributed to the issue discussed in this research.

This new smartphones' design is adopted by multiple companies, and thus new smartphones are being released that use this design without considering the security vulnerability.

This research also proves that communication between the AP and the BPs is vulnerable to attack in Android OS. In addition, it discusses some of the Android security mechanisms that were easily bypassed to accomplish the mission. The paper illustrates some discovered flaws in Android application architecture that allow a break in significant and critical Android operations.

## ACKNOWLEDGMENT

This work was financially supported by Princess Nourah Bint Abdulrahman University.

REFERENCES

[1] C. K. LaDue, V. V. Sapozhnykov, and K. S. Fienberg, "A Data Modem for GSM Voice Channel," *IEEE Transactions on Vehicular Technology*, vol. 57, no. 4, pp. 2205–2218, Jul. 2008.

[2] M. Rashidi, A. Sayadiyan, and P. Mowlaee, "A Harmonic Approach to Data Transmission over GSM Voice Channel," in *3rd International Conference on Information and Communication Technologies: From Theory to Applications, 2008. ICTTA 2008*, 2008, pp. 1–4.

[3] A. Dhananjay, A. Sharma, M. Paik, J. Chen, T. K. Kuppusamy, J. Li, and L. Subramanian, "Hermes: Data Transmission over Unknown Voice Channels," in *Proceedings of the Sixteenth Annual International Conference on Mobile Computing and Networking*, New York, NY, USA, 2010, pp. 113–124.

[4] R. Kratsas. (2012). Unleashing the Audio Potential of SmartphonesMixed Signal Audio Products. Cirrus Logic, http://www.cirrus.com/en/pubs/whitePaper/smartphones_wp.pdf

[5] Francisco Cheng, "Why Smartphones Are Smarter All One Processor," Jun. 2013.

[6] B. W. Lampson, "A Note on the Confinement Problem," *Commun. ACM*, vol. 16, no. 10, pp. 613–615, Oct. 1973.

[7] "National Institute of Standards and Technology-Trusted Computer System Evaluation Criteria." Aug-1983.

[8] R. A. Kemmerer, "A Practical Approach to Identifying Storage and Timing Channels: Twenty Years Later," in *Proceedings of the 18th Annual Computer Security Applications Conference*, Washington, DC, USA, 2002, p. 109–.

[9] S. J. Murdoch and S. Lewis, "Embedding Covert Channels into TCP/IP," in *Information Hiding*, M. Barni, J. Herrera-Joancomartí, S. Katzenbeisser, and F. Pérez-González, Eds. Springer Berlin Heidelberg, 2005, pp. 247–261.

[10] M. Bauer, "New Covert Channels in HTTP: Adding Unwitting Web Browsers to Anonymity Sets," in *In Proceedings of the Workshop on Privacy in the Electronic Society (WPES 2003*, 2003, pp. 72–78.

[11] T. Takahashi and W. Lee, "An assessment of VoIP covert channel threats," in *Third International Conference on Security and Privacy in Communications Networks and the Workshops, 2007. SecureComm 2007*, 2007, pp. 371–380.

[12] N. B. Lucena, J. Pease, P. Yadollahpour, and S. J. Chapin, "Syntax and Semantics-Preserving Application-Layer Protocol Steganography," in *Information Hiding*, J. Fridrich, Ed. Springer Berlin Heidelberg, 2005, pp. 164–179.

[13] M. Z. Rafique, M. K. Khan, K. Alghatbar, and M. Farooq, "Embedding High Capacity Covert Channels in Short Message Service (SMS)," in *Secure and Trust Computing, Data Management and Applications*, J. J. Park, J. Lopez, S.-S. Yeo, T. Shon, and D. Taniar, Eds. Springer Berlin Heidelberg, 2011, pp. 1–10.

[14] K. Papapanagiotou, E. Kellinis, G. F. Marias, and P. Georgiadis, "Alternatives for Multimedia Messaging System Steganography," in *Computational Intelligence and Security*, Y. Hao, J. Liu, Y.-P. Wang, Y. Cheung, H. Yin, L. Jiao, J. Ma, and Y.-C. Jiao, Eds. Springer Berlin Heidelberg, 2005, pp. 589–596.